# WaveMixings.jl: a Julia package for performing on-the-fly time-resolved nonlinear electronic spectra from quasi-classical trajectories

Luis Vasquez and Maxim F. Gelin*

*School of Science, Hangzhou Dianzi University, Hangzhou 310018, China*

Sebastian Pios and Lipeng Chen

*Zhejiang Laboratory, Hangzhou 311100, China*

Zhenggang Lan

*Key Laboratory of Theoretical Chemistry of Environment, Ministry of Education and Guangdong Provincial Key Laboratory of Chemical Pollution and Environmental Safety; School of Environment, South China Normal University, Guangzhou 510006, P. R. China*

Wolfgang Domcke

*Department of Chemistry, Technical University of Munich, D-85747 Garching, Germany*

(Dated: 4th September 2025)

We present an efficient numerical implementation of the quasi-classical doorway-window approximation, specifically designed for on-the-fly simulations of time-resolved nonlinear spectroscopic signals in the Julia package `WaveMixings.jl`. The package contains modules that facilitate standard tasks such as input/output data handling, data filtering, and post-processing, among others. `WaveMixings.jl` includes implementations of various methods developed by the authors, including integral and dispersed transient absorption pump-probe signals and two-dimensional spectra By developing `WaveMixings.jl` we aim to create a versatile platform to perform simulations and develop methodologies within the quasi-classical doorway-window approximation framework.

## I. INTRODUCTION

In recent decades, the study of photochemical reactions at femtosecond time scales has gained significant traction.[1–3] This is in part due to significant progress in the generation of pulsed radiation and the detection of spectroscopic signals across various spectral ranges (*e.g.* visible, ultraviolet (UV), and soft X-ray). Experimentally, nonlinear femtosecond spectroscopy techniques study the time-resolved ultrafast dynamics and nonlinear optical properties of matter (molecules, solids, and biological systems) by exploiting the interaction of ultrashort laser pulses with matter. Common techniques include transient absorption (TA) pump-probe (PP), spectroscopy, transient grating (TG) spectroscopy, and two-dimensional (2D) electronic spectroscopy. 2D electronic spectroscopy is a powerful method that offers detailed spectroscopic insights in two dual frequency dimensions. Experimentally, this is achieved by replacing the pump pulse of PP spectroscopy by a pair of phase-coherent excitation pulses. On the theoretical side, the conventional framework for describing nonlinear optical signals is grounded in perturbation theory applied to the radiation-matter interaction.[4] Alternatively, non-perturbative approaches (not relying in perturbation theory on the radiation-matter interaction) have been devised to describe time-resolved nonlinear signals. These methods incorporate interactions with laser pulses directly into the quantum mechanical equations of motion, specifically the time-dependent Schrödinger equation for isolated molecules or quantum master equations for chromophores in condensed-phase environments.[5]

Simulating the dynamics at conical intersections, which are common in photochemical dynamics reactions,[6–8] poses significant challenges due to strong anharmonicities and non-Born-Oppenheimer effects. For models involving few electronic states and few nuclear degrees of freedom, computationally intensive methods such as the multiconfiguration time-dependent Hartree,[9,10] matrix-product-state,[11–13] or tensor-train network methods[14,15] offer suitable solutions, providing qualitative and quantitative insights. Nonetheless, for systems comprising numerous nuclear degrees of freedom, these aforementioned methods become numerically intense. Consequently, reduced-dimensional models are essential for theoretical modelling in such scenarios. Mixed quantum-classical methods, which treat nuclei classically while electrons are treated quantum mechanically, provide an efficient computational alternative.[16–18] Trajectory surface-hopping (SH) algorithms are among the most widely used methods for simulating nonadiabatic dynamics, in part because they are easy to implement, allow to consider all nuclear degrees of freedom, and nonadiabatic effects can be included at various levels of sophistication.[19]

Nonadiabatic dynamics simulations using *ab-initio* quantum chemical (QM) methods, in conjunction with SH algorithms to simulate TA PP spectra are computationally challenging due to the multitude of single-point electronic-structure calculations required. The package iSPECTRON, for example, offers scripts that use response-function time-dependent density functional theory (TD-DFT) to simulate TA PP and 2D spectra. Depending on the specific requirements, multi-reference electronic-structure methods also can be used.[20,21] Another commonly used package is COBRAMM. COBRAMM provides scripts to automate TA PP simulations at the TD-DFT and QM/molecular mechanics (MM) levels of theory, which allows the simulation of solvent environments.[19] It is worth noting that recent versions of COBRAMM are capable of combining single- and multi-reference method approaches.[22] Octopus is a package for the simulation of TA PP signals with real-time TD-DFT, with the advantage that periodic systems can be treated.[23] The ezSpectra suite (ezFCF and ezDyson) can be used to compute one-photon absorption, photoionization/photodetachment, and nonlinear spectroscopies using wave functions methods.[24] This suite has still to be extended to allow the simulation of TA PP signals. The Quantum Dynamics Toolbox (QDT) is a Matlab software package for the simulation of coherent 2D spectroscopy. Although QDT is open source, the Matlab package requires licensing. Additionally, QDT is focused on model simulations, therefore lacking the interfacing with *ab initio* electronic-structure calculations at the present stage.[25]

The doorway-window (DW) approximation for TA PP spectra was introduced by Yan, Fried and Mukamel in the framework of the third-order response-function formalism.[4,26–28] The basic requirement of the DW approximation is the assumption of non-overlapping pump and probe pulses. Moreover, the pulses have to be short in comparison with the characteristic time scale of the molecular system.

The DW approximation was employed to calculate time-resolved fluorescence spectra,[29–34] femtosecond coherence spectra,[35] three-pulse photon-echo signals,[36,37] pump-dump-probe signals,[38] as well as strong-pulse TA PP,[39–41] and electronic two-dimensional (2D)[42,43] signals of a number of model systems. With the extension of the DW approximation to higher-lying excited electronic states[44–46], it became possible to simulate the contribution of excited-state absorption to TA PP and other nonlinear signals. The DW methodology was also used for the efficient evaluation of nonlinear optical response functions.[47–51]

The DW approximation can be numerically realized in various ways. It was already suggested in Refs. 26–28 to combine quasi-classical or semiclassical trajectory methods with the DW framework. This line of research was further developed in Refs. 52–59 Bonačić-Koutecký, Mitrić, and their colleagues were the first who interfaced a variant of the DW methodology with on-the-fly quasi-classical trajectory simulations, and used it for the evaluation time-resolved photoelectron spectra for a number of polyatomic chromophores.[60–62]

The implementation of the DW approximation with a quasi-classical treatment of the nuclear motion and *ab-initio* on-the-fly electronic-structure calculations was recently developed by Gelin and collaborators.[63] In the quasi-classical DW approximation, the finite duration and the spectral shape of the pump and probe pulses are taken into account.[63] This methodology was recently extended to 2D electronic spectra,[64] time-resolved fluorescence spectra,[65] 2D fluorescence-excitation (2D-FLEX) spectra,[66] strong field PP spectra,[67] and laser-polarization-resolved PP spectra.[68] Recently, the fundamental and numerical aspects of the DW methodology have been reviewed.[69] In these works, in-house scripts were developed for quasi-classical dynamics simulations and the computation of various signals, but these scripts were hitherto not documented. This may pose a challenge for other research groups seeking to use the quasi-classical DW approximation for simulations. This lack of accessibility may hinder the adoption and the further development of this theoretical framework.

Here, we present `WaveMixings.jl`, a package providing software for the simulation of femtosecond time-resolved nonlinear signals in the framework of the quasi-classical DW approximation. Specifically, integral TA PP,[63] dispersed TA PP,[63] 2D,[64] 2D fluorescence-excitation,[66] strong field TA PP, and time-resolved fluorescence signals are considered.[70] `WaveMixings.jl` offers a framework that enables users to simulate these signals from quasi-classical trajectory methods (SH, Ehrenfest, mapping approach and similar) with convenient input options, such as electronic energies and transition dipole moments along trajectories. Users simply need to input a few lines of code to perform the calculations of spectra and can use additional functions to display the results. The source code, along with tutorials, is available as open-access software.[71,72]

We demonstrate the features and capabilities of `WaveMixings.jl` for the prototypical molecule pyrazine. Our objective is to generate spectroscopic signals of pyrazine from quasi-classical DW approximation simulations, which were previously generated using in-house scripts. All computations with `WaveMixings.jl` were conducted on a laptop, with the aim of showcasing the efficiency of the code. Given that the code is open-source, we anticipate that it will facilitate future quasi-classical trajectory simulations of time-resolved nonlinear spectra and further method development by other researchers.

The remainder of this article is structured as follows.

The molecular Hamiltonian used in the DW approximation is described in Section II. Section III introduces the quasi-classical DW approximation framework and the underlying mathematical description for estimating five types of signals (integral TA PP, dispersed TA PP, 2D, 2D-FLEX, strong-pulse and time-resolved fluorescence) are introduced. Section IV summarises the simulation protocols of the different signals. Section V provides detailed technical information about the `WaveMixings.jl` package, including its key features and structure. In Section VI, examples of these signals for the prototypical pyrazine molecule are briefly discussed. In Section VII we summarise our findings and discuss potential applications of `WaveMixings.jl` beyond the simulation of time-resolved signals.

## II. MOLECULAR HAMILTONIAN

The simulations of the nonadiabatic dynamics and time-resolved nonlinear signals are based on the Born-Huang Hamiltonian

$$H_{nm} = [K_N(\boldsymbol{R}, \boldsymbol{P}) + V_n(\boldsymbol{R})]\delta_{nm} - \Lambda_{nm}(\boldsymbol{R}, \boldsymbol{P}). \tag{1}$$

Here $n = 0, 1, 2, ...$ numbers adiabatic electronic states, $\boldsymbol{R}$ and $\boldsymbol{P}$ are nuclear coordinates and momenta, $K_N(\boldsymbol{R}, \boldsymbol{P})$ is the kinetic energy operator of the nuclei, the $V_n(\boldsymbol{R})$ are the adiabatic electronic potential energy functions and the $\Lambda_{nm}(\boldsymbol{R}, \boldsymbol{P})$ are the nonadiabatic coupling (NAC) operators.

Usually, laser pulses are tuned into resonance with lower-lying singlet excited states ($n \geq 1$) which are bright from the electronic ground state ($n = 0$). It is thus appropriate to partition the electronic states $\{n\}$ into three manifolds, $\{0\}, \{I\}, \{II\}$.[5] Manifold $\{0\}$ contains just the electronic ground state. Manifold $\{I\}$ contains the bright electronic states which can be accessed by the pump pulse from the ground state, as well as other electronic states intramolecularly coupled to them; manifold $\{II\}$ comprises those electronic states which can be probed by laser pulses from manifold $\{I\}$. It is thus convenient to rewrite the Hamiltonian of Equation (1) by partitioning it in terms of the manifolds $\{0\}, \{I\},$ and $\{II\}$.

$$H = \begin{pmatrix} H_0 & 0 & 0 \\ 0 & H_\mathrm{I} & 0 \\ 0 & 0 & H_\mathrm{II} \end{pmatrix} \tag{2}$$

where $H_k$ is the molecular Hamiltonian in manifold $k$ ($k = 0, \mathrm{I}, \mathrm{II}$). Thereby, NACs between manifolds $\{0\}$ and $\{\mathrm{I}\}$, $\{\mathrm{I}\}$ and $\{\mathrm{II}\}$, $\{0\}$ and $\{\mathrm{II}\}$ are neglected. On the other hand, NACs within manifold $\{\mathrm{I}\}$ are fully taken into account. NACs within manifold $\{\mathrm{II}\}$ can be ignored to a good approximation, since no trajectories need to be propagated in manifold $\{\mathrm{II}\}$ for the evaluation of the third-order polarization.

Downward and upward TDM operators have the corresponding block structure

$$\boldsymbol{\mu}^\downarrow = \begin{pmatrix} 0 & \boldsymbol{\mu}_{0,\mathrm{I}} & 0 \\ 0 & 0 & \boldsymbol{\mu}_{\mathrm{I},\mathrm{II}} \\ 0 & 0 & 0 \end{pmatrix}, \quad \boldsymbol{\mu}^\uparrow = \begin{pmatrix} 0 & 0 & 0 \\ \boldsymbol{\mu}_{\mathrm{I},0} & 0 & 0 \\ 0 & \boldsymbol{\mu}_{\mathrm{II},\mathrm{I}} & 0 \end{pmatrix} \tag{3}$$

where $\boldsymbol{\mu}_{\mathrm{I},0}$ is responsible for the dipole transitions between the ground state and the states of manifold $\{\mathrm{I}\}$, etc. Electronic transitions between manifolds $\{0\}$ and $\{\mathrm{I}\}$ generate ground-state bleach (GSB) and stimulated emission (SE) contributions to spectroscopic signals, while transitions between all three manifolds $\{0\}$, $\{\mathrm{I}\}$ and $\{\mathrm{II}\}$ in general are responsible for excited-state absorption (ESA).

### III. DOORWAY-WINDOW FRAMEWORK

The DW methodology hinges on several basic assumptions,[69] which are briefly considered below.

(i) Non-overlapping laser pulses. With this assumption, one can recast a general heterodyne-detected four-wave-mixing signal in the DW form,[69]

$$S(\tau, T, \tau_t) \sim \mathrm{Re} \sum_{k=0,\mathrm{I},\mathrm{II}} a_k \mathrm{Tr}[\mathcal{W}_k(\tau_t) \mathcal{L}(T) \mathcal{D}(\tau)]. \tag{4}$$

Here the doorway operator $\mathcal{D}(\tau)$ describes evolution of the system during the coherence time $\tau$ (which corresponds to the delay between the first pair laser pulses); the window operator $\mathcal{W}_k(\tau_t)$ describes evolution of the system during the coherence time $\tau_t$ (which corresponds to the delay between the second pair of laser pulses); the propagator $\mathcal{L}(T)$ governs nonadiabatic field-free evolution of the system during the electronic population time $T$ (which corresponds to the delay between the first and second pair of laser pulses); $k = 0, \mathrm{I}, \mathrm{II}$ correspond to the GSB, SE, and ESA contributions to the signal, respectively; the numerical factors

$$a_k = \begin{bmatrix} 1, & k = 0, \mathrm{I} \\ -1, & k = \mathrm{II} \end{bmatrix},$$

denote the sign of the corresponding term. By convention, GSB and SE contributions carry a positive sign due to enhanced transmission, whereas ESA carries a negative sign because it reduces transmission.

For making Equation (4) suitable for quasi-classical on-the-fly trajectory simulations, we have to resort to several other approximations.

(ii) The nuclear dynamics during each of the laser pulses can be neglected.

(iii) The evolution of electronic coherences during the coherence times $\tau$ and $\tau_t$ is evaluated for fixed nuclei.

(iv) Quantum evolution during the population time $T$ is replaced by the evolution along quasi-classical trajectories, and the quantum trace $\mathrm{Tr}[...]$ is replaced by Monte Carlo sampling of the initial conditions and SH events $\langle ... \rangle_{\mathrm{MC}}$.

The applicability of assumptions (i)-(iv) is discussed in detail in the review.[69] Assumptions (i) and (ii) are rather strong, but may be at least be partially relaxed in future more accurate formulations of the theory. Assumptions (i) and (ii) should be appropriate if short ($\sim 10$ fs) laser pulses are employed for the simulation of the signals. If assumptions (i) and (ii) are satisfied, assumption (iii) is well fulfilled for PP-like spectroscopies (in which the time intervals $\tau$ and $\tau_t$ are controlled by the pulse duration), but may become more restrictive for three-pulse photon echo or 2D spectroscopies (in which $\tau$ and sometimes $\tau_t$ are controlled by electronic dephasing). The validity of assumption (iv) is determined by the validity of the chosen trajectory simulation method and underlying electronis structure methods. The above four assumptions determine the domain of validity the quasi-classical DW approximation.

Following the review,[69] we give below the explicit quasi-classical DW expressions used in `WaveMixings.jl` for the evaluation of spectroscopic signals of interest by the DW methodology.

## A. Integral TA PP signal

The integral TA PP signal is obtained by setting $\tau = \tau_t = 0$ in Equation (4). It is defined as:[5,63,69]

$$S^{int}(T, \omega_{pr}) \sim \mathrm{Re} \sum_{k=0,\mathrm{I},\mathrm{II}} a_k \langle W_k(\omega_{pr}, \boldsymbol{R}(T), \boldsymbol{P}(T)) D(\omega_{pu}, \boldsymbol{R}, \boldsymbol{P}) \rangle_{\mathrm{MC}}. \tag{5}$$

Here $\omega_{pu}$ is the carrier frequency of the pump laser, $\omega_{pr}$ is the carrier frequency of the probe laser, $T$ is the time delay between the pump and probe pulses, $D$ the doorway function, and $W_k$ are the window functions. Explicitly,

$$D(\omega_{pu}, \boldsymbol{R}_g, \boldsymbol{P}_g) = E_{pu}^2(\omega_{pu} - U_{eg}(\boldsymbol{R}_g))|\boldsymbol{\mu}_{ge}(\boldsymbol{R}_g)|^2 \rho_g^{Wig}(\boldsymbol{R}_g, \boldsymbol{P}_g), \tag{6}$$

$$W_0^{int}(\omega_{pr}, \boldsymbol{R}_g(T), \boldsymbol{P}_g(T)) = \sum_{e'} E_{pr}^2(\omega_{pr} - U_{e'g}(\boldsymbol{R}_g(T)))|\boldsymbol{\mu}_{ge'}(\boldsymbol{R}_g(T))|^2, \tag{7}$$

$$W_{\mathrm{I}}^{int}(\omega_{pr}, \boldsymbol{R}_e(T), \boldsymbol{P}_e(T)) = E_{pr}^2(\omega_{pr} - U_{e(T)g}(\boldsymbol{R}_e(T)))|\boldsymbol{\mu}_{ge(T)}(\boldsymbol{R}_e(T))|^2, \tag{8}$$

$$W_{\mathrm{II}}^{int}(\omega_{pr}, \boldsymbol{R}_e(T), \boldsymbol{P}_e(T)) = \sum_f E_{pr}^2(\omega_{pr} - U_{fe(T)}(\boldsymbol{R}_e(T)))|\boldsymbol{\mu}_{e(T)f}(\boldsymbol{R}_e(T))|^2 \tag{9}$$

Hereafter, $g$ labels the electronic ground state; $e$ specifies electronic states of manifold $\{\mathrm{I}\}$, $f$ enumerates electronic states of manifold $\{\mathrm{II}\}$; $\boldsymbol{R}_g(T), \boldsymbol{P}_g(T)$ and $\boldsymbol{R}_e(T), \boldsymbol{P}_e(T)$ denote nuclear positions and momenta in manifolds $\{0\}$ and $\{\mathrm{I}\}$, respectively; $V_g(\boldsymbol{R}), V_e(\boldsymbol{R}), V_f(\boldsymbol{R})$ are the potential energy functions in these states; $U_{eg}(\boldsymbol{R}) = V_e(\boldsymbol{R}) - V_g(\boldsymbol{R})$ and $U_{fe}(\boldsymbol{R}) =$ $V_f(\boldsymbol{R}) - V_e(\boldsymbol{R})$ are the transition frequencies; $\mu_{ge}(\boldsymbol{R})$ and $\mu_{fe}(\boldsymbol{R})$ are the TDMs; $E_{pu}(\omega)$ and $E_{pt}(\omega)$ are Fourier transforms of the temporal envelopes $E_{pu}(t)$ and $E_{pr}(t)$ of the pump and probe pulses; the notation $e(T)$ implies that a trajectory launched at $t = 0$ in a state $e$ can end up at time $t = T$ in another state $e(T) \neq e$; $\rho_g^{Wig}(\boldsymbol{R}, \boldsymbol{P})$ is the vibrational Wigner distribution in the electronic ground state $g$.[73]

In the present variant of `WaveMixings.jl`, we do not explicitly consider polarizations of laser pulses and orientational averaging. Hence, all DW formulas of this section contain TDMs squared. The latter are proportional to the oscillator strengths $F$ of the corresponding transitions divided by the energy differences,

$$|\boldsymbol{\mu}_{ge}(\boldsymbol{R})|^2 = \frac{3F_{ge}(\boldsymbol{R})}{2U_{eg}(\boldsymbol{R})}, \quad |\boldsymbol{\mu}_{ef}(\boldsymbol{R})|^2 = \frac{3F_{ef}(\boldsymbol{R})}{2U_{fe}(\boldsymbol{R})}. \tag{10}$$

In principle, however, the DW methodology is capable of proper treatment of the polarizations of the laser pulses and orientational averaging. This was demonstrated, for example, in Ref. 74 for the example of azobenzene, and in Refs. 75,76 for dendrimers.

## B. Dispersed TA PP signal

Experimentally, the spectrum of a transmitted probe pulse can be dispersed by a spectrometer and recorded as a function of the delay time $T$ and the dispersion frequency $\omega_t$. The dispersed TA PP signal is obtained from Equation (4) by setting $\tau = 0$ and performing a Fourier transform (frequency $\omega_t$) with respect to $\tau_t$:[5,63,69]

$$S^{dis}(T, \omega_t) \sim \operatorname{Re} \sum_{k=0,\mathrm{I},\mathrm{II}} a_k \langle W_k^{\mathrm{dis}}(\omega_{pr}, \omega_t, \boldsymbol{R}(T), \boldsymbol{P}(T)) D(\omega_{pu}, \boldsymbol{R}, \boldsymbol{P}) \rangle_{\mathrm{MC}}. \tag{11}$$

Here the doorway function $D$ is the same as in the integral signal (Equation (6)), while the window functions for the dispersed signal read

$$W_0^{dis}(\omega_{pr}, \omega_t, \boldsymbol{R}_g(T), \boldsymbol{P}_g(T)) = E_{pr}^2(\omega_t - \omega_{pr}) \sum_e \frac{\nu}{\nu^2 + (\omega_t - \omega_{eg}(\boldsymbol{R}_g(T)))^2} |\mu_{ge}(\boldsymbol{R}_g(T))|^2, \tag{12}$$

$$W_\mathrm{I}^{dis}(\omega_{pr}, \omega_t, \boldsymbol{R}_e(T), \boldsymbol{P}_e(T)) = E_{pr}^2(\omega_t - \omega_{pr}) \frac{\nu}{\nu^2 + (\omega_t - \omega_{e(T)g}(\boldsymbol{R}_e(T)))^2} |\mu_{ge(T)}(\boldsymbol{R}_e(T))|^2, \tag{13}$$

$$W_{\text{II}}^{dis}(\omega_{pr}, \omega_t, \boldsymbol{R}_e(T), \boldsymbol{P}_e(T)) = E_{pr}^2(\omega_t - \omega_{pr}) \sum_f \frac{\nu}{\nu^2 + (\omega_t - \omega_{fe(T)}(\boldsymbol{R}_e(T)))^2} |\mu_{e(T)f}(\boldsymbol{R}_e(T))|^2. \tag{14}$$

In the above formulas, $\nu$ is the electronic dephasing rate. It is a technical parameter specifying the line shape of $I^{dis}(T, \omega_t)$ in the frequency domain.

## C. Time-resolved fluorescence signal

In the time-resolved fluorescence (TRF) experiment, fluorescence from the chromophore is mixed with an up-conversion pulse in a nonlinear medium. The quasi-classical DW expression for the TRF spectrum was derived by Xu *et al.*[65] Apart from the frequency-dependent prefactor, the TRF spectrum is identical to the SE contribution to the integral TA PP signal:[5,65,69]

$$S^{TRF}(T, \omega) \sim \omega_{pu} \omega^3 \langle W_{\text{I}}^{\text{TFR}}(\boldsymbol{R}(T), \boldsymbol{P}(T); \omega) D(\boldsymbol{R}_g, \boldsymbol{P}_g) \rangle_{\text{MC}}. \tag{15}$$

Here $\omega$ denotes frequency of fluorescence photons emitted at time $T$, $D$ is given by Equation (6), while $W_{\text{I}}^{\text{TFR}}$ is obtained from Equation (8) by replacing $\omega_{pr} \to \omega$. In this case, $E_{pr}(\omega)$ defines the profile of the up-conversion pulse in the frequency domain.

## D. Two-Dimensional (2D) signal

The electronic 2D spectrum is obtained by Fourier transforming Equation (4) with respect to $\tau$ (frequency $\omega_\tau$) and $\tau_t$ (frequency $\omega_t$):[5,64,69]

$$S_\alpha^{2D}(\omega_\tau, T, \omega_t) \sim \text{Re} \sum_{k=0,\text{I},\text{II}} a_k \langle W_k(\boldsymbol{R}(T), \boldsymbol{P}(T); \omega_t) D_\alpha(\boldsymbol{R}, \boldsymbol{P}; \omega_\tau) \rangle_{\text{MC}}. \tag{16}$$

Here $\alpha = \text{NR}$, $\xi_\alpha = 1$ correspond to the non-rephasing spectrum, $\alpha = \text{R}$, $\xi_\alpha = -1$ correspond to the rephasing spectrum, the doorway function reads

$$D_\alpha^{2D}(\omega_\tau, \boldsymbol{R}_g, \boldsymbol{P}_g) = \rho_g^{Wig}(\boldsymbol{R}_g, \boldsymbol{P}_g) \frac{E_{pu}^2(\omega_\tau - \omega_{pu})|\mu_{ge}(\boldsymbol{R}_g)|^2}{\nu + i\xi_\alpha(U_{eg}(\boldsymbol{R}_g) - \omega_\tau)}, \tag{17}$$

while the window functions are

$$W_0^{2D}(\omega_t, \boldsymbol{R}_g(T), \boldsymbol{P}_g(T)) = \sum_{e'} \frac{E_{pr}^2(\omega_t - \omega_{pr})|\mu_{ge'}(\boldsymbol{R}_g(T))|^2}{\nu + i(U_{e'g}(\boldsymbol{R}_g(T)) - \omega_t)}, \tag{18}$$

$$W_{\text{I}}^{2D}(\omega_t, \boldsymbol{R}_{e(T)}(T), \boldsymbol{P}_{e(T)}(T)) = \frac{E_{pr}^2(\omega_t - \omega_{pr})|\mu_{ge(T)}(\boldsymbol{R}_{e(T)}(T))|^2}{\nu + i(U_{e(T)g}(\boldsymbol{R}_{e(T)}(T)) - \omega_t)}, \tag{19}$$

$$W_{\text{II}}^{2D}(\omega_t, \boldsymbol{R}_{e(T)}(T), \boldsymbol{P}_{e(T)}(T)) = \sum_f \frac{E_{pr}^2(\omega_t - \omega_{pr})|\mu_{e(T)f}(\boldsymbol{R}_{e(T)}(T))|^2}{\nu + i(U_{fe(T)}(\boldsymbol{R}_{e(T)}(T)) - \omega_t)}. \tag{20}$$

Both the doorway and window functions of the 2D signal are complex-valued and depend on the electronic dephasing rate $\nu$ introduced in Section III B.

### E. 2D-FLEX signal

2D-FLEX is a novel nonlinear femtosecond fluorescence-detection technique that merges the spectral resolution of excitation and detection of 2D electronic spectroscopy with the selective detection of excited state dynamics inherent to TRF.[77] As a consequence, the 2D-FLEX spectrum enables the detailed monitoring of molecular wavepacket motion in manifold {I} (SE contribution) with high temporal and spectral resolution, whilst excluding GSB and ESA contributions. Pios *et al.* developed the DW methodology for calculating the 2D-FLEX spectrum:[66]

$$S^{2\text{DF}}(\omega_\tau, T, \omega_t) \sim$$
$$\left\langle W_{\text{I}}^{2\text{DF}}(\omega_t, \boldsymbol{R}_e(T), \boldsymbol{P}_e(T)) D_{\text{I}}^{2\text{DF}}(\omega_\tau, \boldsymbol{R}_g, \boldsymbol{P}_g) \right\rangle. \tag{21}$$

where

$$D^{2\text{DF}}(\omega_\tau, \boldsymbol{R}_g, \boldsymbol{P}_g; e) =$$
$$\rho_g^{Wig}(\boldsymbol{R}_g, \boldsymbol{P}_g) E_\nu^2(U_{eg}(\boldsymbol{R}_g) - \omega_\tau) E_{pu}^2(\omega_\tau - \omega_{pu}) |\mu_{ge}(\boldsymbol{R}_g)|^2, \tag{22}$$

and

$$W^{2\text{DF}}(\omega_t, \boldsymbol{R}_e(T), \boldsymbol{P}_e(T); e(T)) =$$
$$\tilde{E}_{pr}^2(\omega_t - U_{e(T)g}(\boldsymbol{R}_e(T))) |\mu_{e(T)g}(\boldsymbol{R}_e(T))|^2. \tag{23}$$

In the above formulas, we explicitly assume that electronic dephasing is described by a Gaussian function $\exp\{-(t/\tau_\nu)^2\}$ with a characteristic time $\tau_\nu$ (inhomogeneous broadening), the probe pulse of duration $\tau_{pr}$ has a Gaussian envelope ($E_{pr}(t) = \exp\{-(t/\tau_{pr})^2\}$, $E_{pr}(\omega) = \exp\{-(\omega\tau_{pr})^2/4\}$) and

$$\tilde{E}_{pr}(\omega) = \exp\left(-\frac{\omega^2}{4/\tau_{pr}^2 + 8/\tau_\nu^2}\right). \tag{24}$$

## F. Strong field TA PP signal

In the weak-field limit, the doorway and window functions are proportional to the sintensities of the pump and probe pulses. In other words, $D \sim \lambda_{pu}^2$ and $W \sim \lambda_{pr}^2$, where $\lambda_{pu}$ and $\lambda_{pr}$ are amplitudes of the pump and probe pulses, respectively. Beyond the weak-field limit, this scaling breaks down.[40,41] Nevertheless, if we neglect population contributions stemming from manifold {II} and assume rectangular pulse envelopes,

$$E_j(t) = \begin{cases} 1, \text{ if } -\tau_j/2 \leq t \leq \tau_j/2 \\ 0, \quad\quad\quad\quad \text{otherwise} \end{cases} \quad (25)$$

($j = pu, pr$), the strong field TA PP spectra can be evaluated with the semi-classical DW method as follows:[67]

$$S^{str}(T, \omega_{pr}) \sim \text{Re} \sum_{k=0,\text{I,II}} a_k \langle W_k^{str}(\omega_{pr}\boldsymbol{R}(T), \boldsymbol{P}(T)) D^{str}(;\omega_{pu}\boldsymbol{R}, \boldsymbol{P}) \rangle. \quad (26)$$

Here the doorway functions is

$$D^{str}(\omega_{pu}, \boldsymbol{R}_g, \boldsymbol{P}_g) = \sum_{e'} \frac{\lambda_{pu}^2 |\mu_{ge}(\boldsymbol{R}_g)|^2}{\Omega_{eg}^2(\boldsymbol{R}_g)} (1 - \cos(\Omega_{eg}(\boldsymbol{R}_g)\tau_{pu})) \rho_g^{\text{Wig}}(\boldsymbol{R}_g, \boldsymbol{P}_g), \quad (27)$$

where

$$\Omega_{eg}(\boldsymbol{R}_\text{g}) = \sqrt{(U_{eg}(\boldsymbol{R}_\text{g}) - \omega_{pu})^2 + 4\lambda_{pu}^2|\mu_{ge}(\boldsymbol{R}_\text{g})|^2}. \quad (28)$$

is the Rabi frequency. The window functions are determined by similar in structure expressions

$$W_0^{str}(\omega_{pr}, \boldsymbol{R}_g(T), \boldsymbol{P}_g(T)) = \sum_{e'} \frac{\lambda_{pr}^2 |\mu_{ge}(\boldsymbol{R}_g(T))|^2}{\Omega_{eg}^2(0, \boldsymbol{R}_g(T))} (1 - \cos(\Omega_{eg}(0, \boldsymbol{R}_g(T))\tau_{pu})), \quad (29)$$

$$W_\text{I}^{str}(\omega_{pr}, \boldsymbol{R}_{e(T)}(T), \boldsymbol{P}_{e(T)}(T)) = \frac{\lambda_{pr}^2 |\mu_{ge(T)}(\boldsymbol{R}_e(T))|^2}{\Omega_{e(T)g}^2(0, \boldsymbol{R}_e(T))} (1 - \cos(\Omega_{e(T)g}(0, \boldsymbol{R}_e(T))\tau_{pu})), \quad (30)$$

$$W_\text{II}^{str}(\omega_{pr}, \boldsymbol{R}_{e(T)}(T), \boldsymbol{P}_{e(T)}(T)) = \sum_f \frac{\lambda_{pr}^2 |\mu_{fe(T)}(\boldsymbol{R}_e(T))|^2}{\Omega_{fe(T)}^2(0, \boldsymbol{R}_e(T))} (1 - \cos(\Omega_{e(T)f}(0, \boldsymbol{R}_e(T))\tau_{pu})), \quad (31)$$

where the respective Rabi frequencies of the window functions are

$$\Omega_{eg}(0, \boldsymbol{R}_g(T)) = \sqrt{(U_{eg}(\boldsymbol{R}_g(T)) - \omega_{pr})^2 + 4\lambda_{pr}^2 |\boldsymbol{\mu}_{ge}(\boldsymbol{R}_g(T))|^2}, \tag{32}$$

$$\Omega_{eg}(\mathrm{I}, \boldsymbol{R}_g(T)) = \sqrt{(U_{e(T)g}(\boldsymbol{R}_e(T)) - \omega_{pr})^2 + 4\lambda_{pr}^2 |\boldsymbol{\mu}_{ge(T)}(\boldsymbol{R}_e(T))|^2}, \tag{33}$$

$$\Omega_{eg}(\mathrm{II}, \boldsymbol{R}_g(T)) = \sqrt{(U_{e(T)f}(\boldsymbol{R}_e(T)) - \omega_{pr})^2 + 4\lambda_{pr}^2 |\boldsymbol{\mu}_{ge(T)}(\boldsymbol{R}_e(T))|^2}. \tag{34}$$

In the weak-field limit, the strong-field formulas reduce to their weak-field counterparts of Section III A for rectangular pulse envelopes.

## IV. SIMULATION PROTOCOLS

The initialization of a classical trajectory requires the specification of the initial nuclear coordinates and momenta $\boldsymbol{R}_g$, $\boldsymbol{P}_g$ in the electronic ground state and the selection of the initially excited state $e$ of manifold $\{\mathrm{I}\}$. $\boldsymbol{R}_g$ and $\boldsymbol{P}_g$ are sampled from the Wigner distribution $\rho_g^{Wig}(\boldsymbol{R}_g, \boldsymbol{P}_g)$ of the electronic ground state. The initial electronic state excited by the pump pulse is sampled from the oscillator strength distribution of the electronic states in manifold $\{\mathrm{I}\}$. This sampling distribution

$$\rho_s(\boldsymbol{R}_g, \boldsymbol{P}_g, e) = |\boldsymbol{\mu}_{ge}(\boldsymbol{R}_g)|^2 \rho_g^{Wig}(\boldsymbol{R}_g, \boldsymbol{P}_g). \tag{35}$$

is the standard choice for the simulation of time-dependent electronic population probabilities with quasi-classical propagation methods. The sampling distribution of Equation (35) is appropriate for very short (sub-fs) pump pulses. The power spectrum of the pump pulse, $E_{pu}^2(\omega_{pu} - U_{eg}(\boldsymbol{R}_g))$, is then included in the evaluation of the doorway function, Equation (6). For longer pulses with a narrower frequency spectrum, it is computationally more efficient to include the power spectrum of the pump pulse in the sampling distribution, using

$$\rho_s(\boldsymbol{R}_g, \boldsymbol{P}_g, e) = E_{pu}^2(\omega_{pu} - U_{eg}(\boldsymbol{R}_g)) |\boldsymbol{\mu}_{ge}(\boldsymbol{R}_g)|^2 \rho_g^{Wig}(\boldsymbol{R}_g, \boldsymbol{P}_g). \tag{36}$$

With any of these choices, the counterpart of the basic DW Equation (4) employing sampling from $\rho_s$ is given by the formula

$$S(\tau, T, \tau_t) \sim \mathrm{Re} \sum_{k=0,\mathrm{I},\mathrm{II}} a_k \langle W_k(\boldsymbol{R}(T), \boldsymbol{P}(T); \tau_t) D^{(s)}(\boldsymbol{R}_g, \boldsymbol{P}_g; \tau) \rangle_{\mathrm{MCs}}. \tag{37}$$

in which $\langle...\rangle_{\text{MCs}}$ means MC sampling of the initial trajectories from $\rho_s$ and

$$D^{(s)}(\boldsymbol{R}_g, \boldsymbol{P}_g; \tau) = \rho_s^{-1}(\boldsymbol{R}_g, \boldsymbol{P}_g, e) D(\boldsymbol{R}_g, \boldsymbol{P}_g; \tau). \tag{38}$$

For example, $\rho_s$ of Equation (35) yields the TA PP doorway function

$$D^{(s)}(\boldsymbol{R}_g, \boldsymbol{P}_g) = E_{pu}^2(\omega_{pu} - U_{eg}(\boldsymbol{R}_g)) \tag{39}$$

If $\rho_s$ of Equation (36) is used, then the TA PP doorway function is constant,

$$D^{(s)}(\boldsymbol{R}_g, \boldsymbol{P}_g) = 1. \tag{40}$$

The computational steps involved in the simulation of any DW signal with classical trajectories can be summarized as follows.

1. Sample an initial condition $\boldsymbol{R}_g$, $\boldsymbol{P}_g$ and $e$ from the phase-space distribution $\rho_s(\boldsymbol{R}_g, \boldsymbol{P}_g, e)$ and calculate the doorway function $D^{(s)}(\boldsymbol{R}_g, \boldsymbol{P}_g; \tau)$ of Equation (38).

2. Propagate the trajectory $\boldsymbol{R}_g(t)$, $\boldsymbol{P}_g(t)$ with the classical ground-state Hamiltonian $H_0$ up to the final time $t_F$ of the simulation.

3. Evaluate $W_0(\boldsymbol{R}_g(T), \boldsymbol{P}_g(T))$ for the desired grid of pulse delay times $T$.

4. Propagate the trajectory $\boldsymbol{R}_e(t)$, $\boldsymbol{P}_e(t)$ with the SH algorithm up to the final time $t_F$ of the simulation.

5. Evaluate $W_{\text{I}}(\boldsymbol{R}_e(T), \boldsymbol{P}_e(T))$ and $W_{\text{II}}(\boldsymbol{R}_e(T), \boldsymbol{P}_e(T))$ for the desired grid of pulse delay times $T$.

6. Repeat these calculations until the MC sampling is converged.

## V. STRUCTURE OF WAVEMIXINGS.JL

This section offers a comprehensive overview of the coding strategy, key functionalites, and structural composition of the `WaveMixings.jl` package,[71] as shown in Figure 1. `WaveMixings.jl` is a package that offers four major functionalitiesi which represent services provided by the package. This approach is possible because Julia, although not strictly an object-oriented programming language (OOP), is a multi-paradigm language that includes

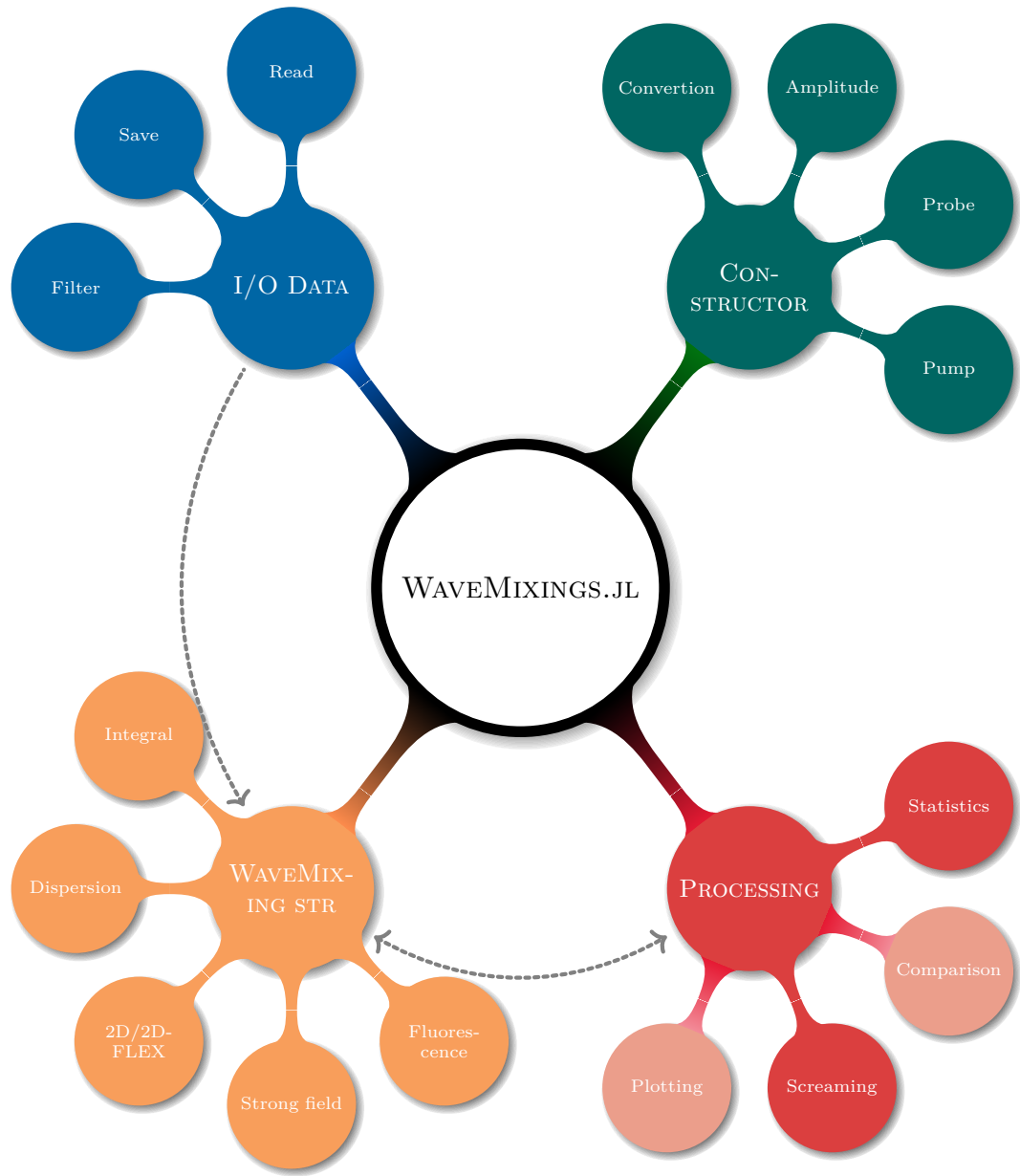

Figure 1: `WaveMixings.jl` internal structure and auxiliary functionalities (data, constructor, WaveMixing structure, and processing). Grey dashed lines indicate direct connection from one functionality to other in direction of the pointing arrow.

OOP patterns and emphasizes multiple dispatch efficiency, allowing for generic code, dynamic type assignment, therefore preventing unintentional OOP issues such as mixing of classes and attaching multi-methods defined in separate files. These specific features of the Julia programming language speed up the development of `WaveMixings.jl` and method implementation for users who are inexperienced in programming. Readers are encouraged to refer to S I for more information on the source code or Section III for details on the theoretical DW framework and mathematical details that serve as the foundation for `WaveMixings.jl`.

The primary purpose of `WaveMixings.jl` is to simulate nonlinear spectroscopic signals within the theoretical framework of the quasi-classical DW approximation. At present, this package is capable of post-processing output files from trajectory surface-hopping simulations using the code ZagHop.[78–80] ZagHop has an output structure (legacy output[78] in S 1 and modern output[79] S 1 to 3) which is similar to those of other trajectory SH software. The output files of other dynamics codes may be adapted to the ZagHop output structure for subsequent data extraction with `WaveMixings.jl` with minimal data post-processing (see examples in Ref. 72). Alternatively, `WaveMixings.jl` users have the option to directly input formatted data into the `energies` and `oscillators` fields of the Julia object `WaveMixing` (see S 4 and 5). For more detailed instructions of input file structure, `WaveMixing` fields format, and other importing tools, readers are encouraged to consult S II.

`WaveMixings.jl` comprises four key functionalites:

**I/O Data:** is the functionality responsible for importing, exporting, and filtering of data. It includes recipes for importing the data files of GSB, SE, and ESA signals (see Section III for more details). At present, `WaveMixings.jl` can handle output files from ZagHop. However, users can directly import post-processed data into `WaveMixings.jl` when such data are available.

**Constructors:** includes functions that transform input parameters or data into atomic units, generate arrays with pump/probe values, and convert axis values for plotting. This functionality is indirectly linked with other `WaveMixings.jl` modules to ensure that all calculations are executed in atomic units.

**Processing:** contains functions to normalise the spectra, perform comparisons when necessary, and plot the results using colour-blind friendly palettes. Given that the spectra might include complex-valued data, this functionality ensures that the spectra are properly normalised and plotted. Additionally, it includes functionalities to identify missing values and to smooth (interpolate) the results for improved visualisation.

**WaveMixing struct (str):** is a Julia struct designed with internal functions to compute time-resolved pump-probe observables such as GSB, SE, and ESA. Each observable is associated with a strongly typed function tailored for the implemented phenomena, e.g. `gsb`, `gsb_strong`, and `gsb_dispersion`.

`WaveMixings.jl` is an open-source package hosted in `codeberg.org`.[71] This allows users to fork, independently develop, or merge code into the package. Entirely written in the Julia programming language, `WaveMixings.jl` delivers high performance that often surpasses Python, R, and Matlab. It is designed to be user-friendly, versatile, and numerically efficient, using Julia's multiple dispatch feature. Every function within `WaveMixings.jl` is strongly typed, similar to C++, enabling the package to take full advantage of multiple dispatch for enhanced optimisation and reducing the likelihood of type errors. Moreover, most `WaveMixings.jl` functions use the Basic Linear Algebra Subprograms (BLAS) from the `LinearAlgebra.jl` package. This ensures these functions are parallelised and well-suited for use on desktop environments, laptops, and high-performance computing infrastructures.

The source code of `WaveMixings.jl` leverages solely the existing packages within Julia's standard library –`LinearAlgebra.jl` and `DelimitedFiles.jl`– enabling calculations directly from a raw Julia installation. All functions within `WaveMixings.jl` are thoroughly documented, accessible interactively through Julia's help system, and additionally presented on the `codeberg` Wiki alongside examples and an API list. Ref. 72 provides two sets of examples that cover technical details for employing electronic structure calculations alongside SH algorithms for simulating GSB, SE, and ESA signals. These examples also include `WaveMixings.jl` scripts for calculating TA PPs and generating 2D plots. These examples serve as templates. For realistic simulations, more trajectories are required to obtain converged signals, see Ref. 63,64 for examples of converged simulations.

We tested `WaveMixings.jl v0.3.5` with Julia versions `1.11.5`, `1.12.0-rc1`, and `1.13.0-DEV` on an i5-12500H Intel platform equipped with 16 GB of RAM. Section VI showcases sample codes and signal examples of signals obtained with `WaveMixings.jl`, whereas S III details code examples for the calculation of all signals are included.

## VI. ILLUSTRATIVE EXAMPLES

We provide examples for the simulation of time-resolved nonlinear signals, namely integral TA PP (Section VI A), dispersed TA PP (Section VI B), TRF (Section VI C), 2D

## A. Integral TA PP signal for pyrazine

This section presents TA PP spectra of pyrazine obtained with `WaveMixings.jl`. The *ab initio* trajectory data are those of Ref. 63 (see supplementary information). The details of the SH simulation calculations are identical to those outlined in Ref. 63. In manifold {I}, four excited states with vertical excitation energies $4.18\,\text{eV}$, $4.83\,\text{eV}$, $5.08\,\text{eV}$ and $5.08\,\text{eV}$ are included. In manifold {II}, 30 electronic states with vertical excitation energies $\leq 10\,\text{eV}$ are included. The `filter_files_size` function from `WaveMixings.jl` I/O Data functionality was used to filter and select only those SH trajectories that were successfully completed (591 out of 600 in our case).

For the simulation of the TA PP signal of pyrazine, aside from the trajectory output logs, `WaveMixings.jl` requires the durations of the pump and probe pulses ($\tau = 5\,\text{fs}$), the pump carrier frequency ($\omega_{\text{pu}} = 5.2\,\text{eV}$ in steps of $0.02\,\text{eV}$), and the range of the probe carrier frequencies ($\omega_{\text{pr}} = 2.0\,\text{eV}$ to $6.0\,\text{eV}$). By default, `WaveMixings.jl` assumes that the temporal shape of pulses is Gaussian. In addition, `WaveMixings.jl` contains implementations for asymmetric squared hyperbolic secant-, squared hyperbolic secant-, and Lorentzian-shaped pulses that are accessible trough the field `method` of the `WaveMixing` Julia struct. See a code example for the GSB contribution calculation in Listing 1 and detailed code explanation in S III. Because `WaveMixings.jl` performs all computations in atomic units, the constructor functionality includes functions to convert units or initialize variables (such as pump carrier frequency, pulse duration time, *etc.*). Specifically, for the pump frequency, the constructor creates a Julia Array with a predetermined number of points within an energy range – e.g. `make_pump_freq(2.0, 6.0, 200)` in this simulation. Evaluating the TA PP signal of pyrazine with the input from 591 trajectory logs (200 GSB, 191 SE, and 200 ESA) takes $128\,\text{s}$.

Our implementation of the quasi-classical DW approximation within `WaveMixings.jl` reproduces the TA PP signals of Gelin *et al.* for pyrazine as depicted Figure 2.[63] The GSB signal exhibits a peak near $\omega_{\text{pr}} \approx 5.2\,\text{eV}$ which oscillates in energy and intensity with a period of approximately $\approx 33\,\text{fs}$. Additionally, there is a second, less prominent contribution at $\omega_{\text{pu}} \approx 4.1\,\text{eV}$, representing the $^1\text{B}_{3u}(n\pi^*)$ state of pyrazine which is weakly allowed due to vibronic coupling. The SE signal plot exhibits the ultrafast radiationless decay of the $^1\pi\pi^*$ state to the $^1\pi^*$ states on a time scale of about $\approx 20\,\text{fs}$. The SE signal also shows features of the dynamics of the vibrationally hot $^1B_{3u}$ and $^1A_u$ state with intensity revivals at $85\,\text{fs}$ and $150\,\text{fs}$.

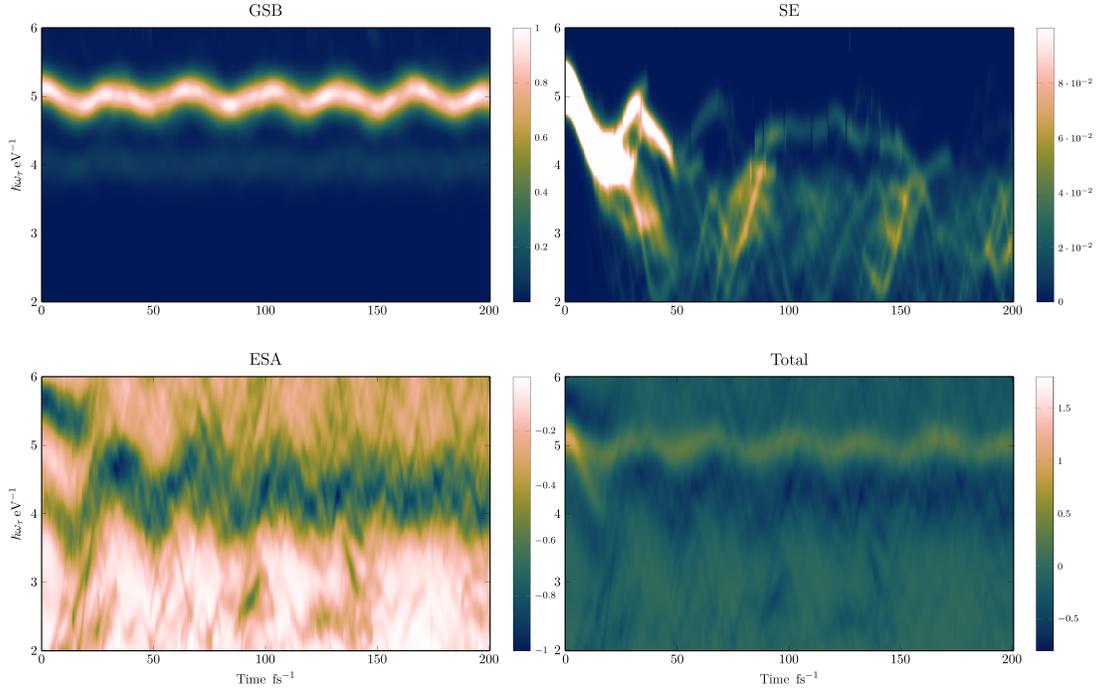

Figure 2: TA PP signals (GSB, SE, and ESA) of pyrazine and the total signal. Input values: pulse duration $(\tau) = 5\,\text{fs}$, pump frequency $(\omega_{\text{pu}}) = 5.2\,\text{eV}$. The signal has been evaluated for probe carrier frequencies in the range from $2.0\,\text{eV}$ to $6.0\,\text{eV}$ in steps of $0.02\,\text{eV}$.

The ESA signal is complex due to the multitude of states in manifold {II} involved. Nevertheless, the ESA spectrum exhibits a distinct maximum at $\omega_{\text{pr}} \approx 5.8\,\text{eV}$ that appears to oscillate in phase alongside the GSB signal. In contrast to the GSB signal, the oscillations of the ESA signal are damped on a time scale of several hundred femtoseconds. Lastly, in Figure 2, we show the total signal, which is the sum of GSB, SE, and ESA components. The total signal is dominated by the GSB signal. Although the SE signal is comparatively weak, its decay within the initial $20\,\text{fs}$ is visible in the total signal.

```julia
1  # WaveMixings parameters of the signal
2  pump_frequ = make_pump_freq(5.2)
3  pulse_time = make_pulse_duration(5.0)
4  probe_freq = make_probe_freq(2.0, 6.0, 200)
5  # Read file with data from SH calculation
6  tmp_GS = extract_data_gs("001_good_gsb.log")
7  gsb(pump_frequ, pulse_time, tmp_GS)
8  gsb(probe_freq, pulse_time, tmp_GS)
9  # Create the Doorway-window and store the result
10 # in the "spectral_signal" field.
11 make_door_window(tmp_GS)
```

Listing 1: Julia code example illustrating the calculation of the GSB component of the integral TA PP signal. The code for the SE and ESA components is similar, see S 6 for the SE example).

## B. Dispersed TA PP signal of pyrazine

In `WaveMixings.jl`, dispersed signals are designated by appending "dispersion" to the signal name, for instance, `gsb_dispersion`. The argument `dephase` specifies the optical dephasing parameter ($v$). See S 7 for a code example of the GSB contribution to the dispersed TA PP signal. Figure 3 presents the dispersed TA PP signals for pyrazine. The carrier frequency of the pump pulse is $5.2\,\text{eV}$. The signals are rather narrow in the frequency domain due to the limited spectral width of the $5\,\text{fs}$ probe pulse.

## C. TRF signal of pyrazine

The TRF signal shown in Figure 4 resembles the SE part of the integral TA PP signal in Figure 2. An example code and detailed code explanation for the simulation of the GSB contribution can be found in S 8. However, unlike the SE part in the integral TA PP signal, the TRF signal lacks intensity below $3.0\,\text{eV}$ and exhibits significantly less pronounced peaks at $T \approx 85\,\text{fs}$ and $150\,\text{fs}$.

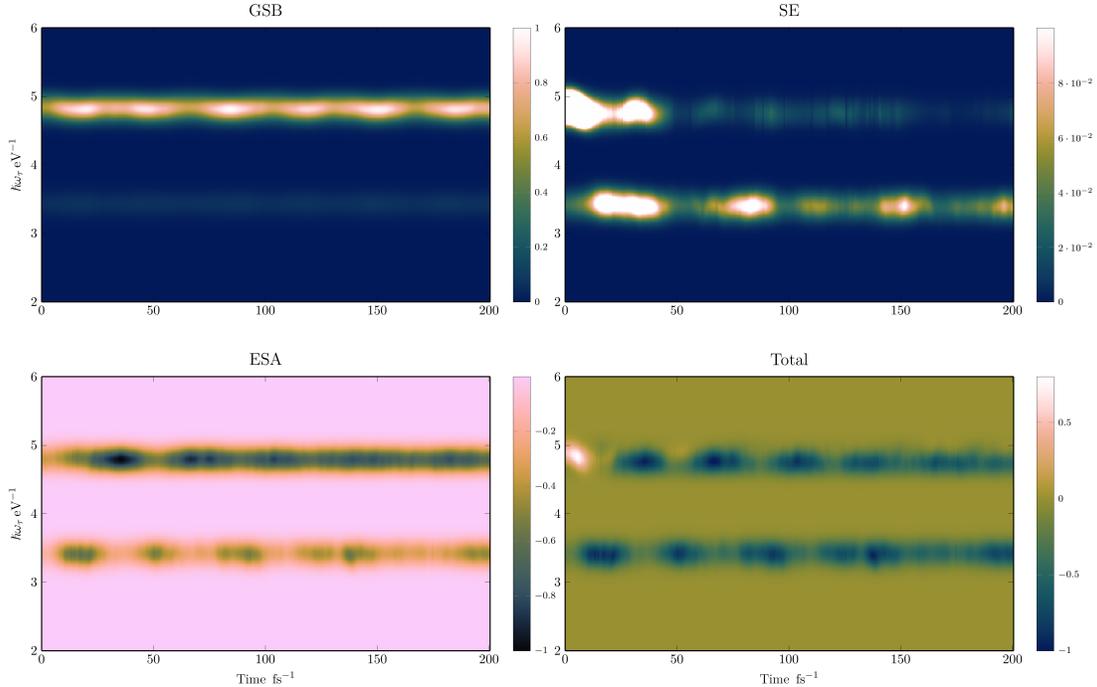

Figure 3: Dispersed TA PPP signals (GSB, SE, and ESA) of pyrazine and the total signal. `WaveMixings.jl` input values are: pulse duration ($\tau$) = 5 fs, dephasing parameter ($\upsilon$) = 0.001 eV. The dispersed probe frequency varies from 2.0 eV to 6.0 eV in steps of 0.02 eV.

### D. 2D TA signal of pyrazine

In `WaveMixings.jl`, we have implemented the 2D TA signal as described in reference 64, including the two phase-matching directions - rephasing (`r2d`) and nonrephasing (`nr2d`)- for comparative analysis. For the GSB contribution to the 2D signal (and similarly for SE and ESA), the Julia functions are: `gsb_door_r2d`, `gsb_door_nr2d`, and `gsb_wind_2d` (see a code example in Listing 2 and detailed code explanation in S III).

We evaluated `nr2d` 2D TA plots for GSB, SE, and ESA contributions, alongside the total signal at a waiting time of 60 fs. Figure 5 depicts the 2D TA of pyrazine at 60 fs, where transitions from the $^1A_u$ state dominate the ESA signal, as noted by Huang *et al.*[64]

### E. 2D-FLEX signal of pyrazine

Figure 6 showcases the simulated 2D-FLEX signal of pyrazine at a waiting time of 60 fs. Notably, Figure 6 reveals that the spectra exhibit prominent doublets at frequencies $\omega_{\mathrm{t}} = 4.5\,\mathrm{eV}$ and $\omega_{\mathrm{t}} = 3.2\,\mathrm{eV}$. The doublets are surrounded by less intense peaks along the excitation frequency of the bright $^1\mathrm{B}_{2u}$ state. The aforementioned peaks emerge from lower vibronic levels of the $^1\mathrm{B}_{2u}$ state and higher vibronic levels of the $^1\mathrm{B}_{3u}/^1\mathrm{A}_u$ states.

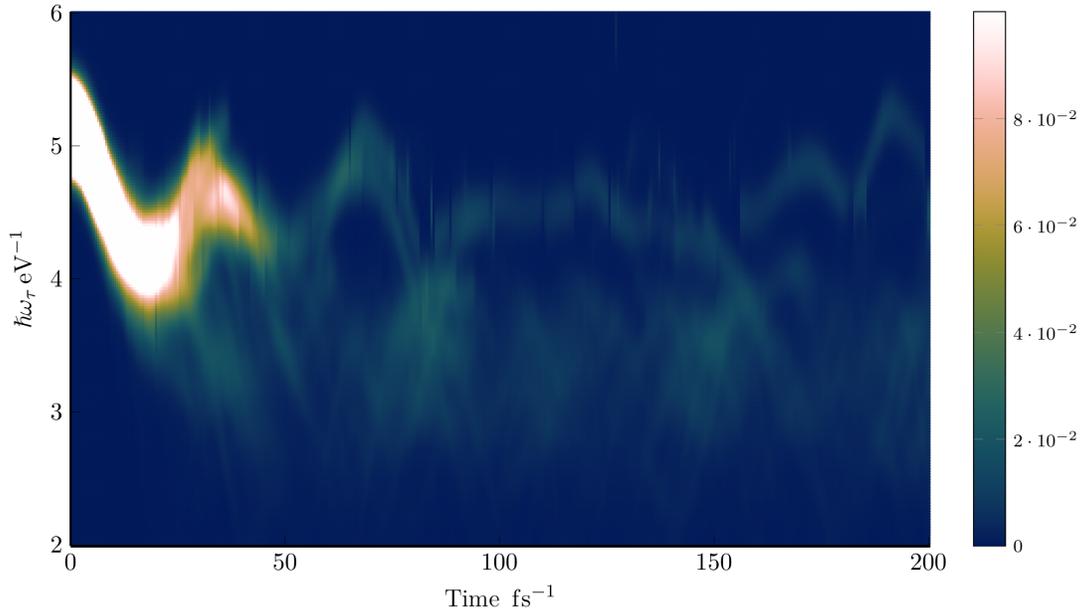

Figure 4: TRF signal of pyrazine.. `WaveMixings.jl` input parameters are: pulse duration ($\tau$) = 5 fs for pump and probe pulses, pump frequency ($\omega_{\mathrm{pu}}$) = 5.2 eV, probe frequency ($\omega_{\mathrm{pr}}$) covers the range from 2.0 eV to 6.0 eV in steps of 0.02 eV.

## VII. CONCLUSIONS AND OUTLOOK

`WaveMixings.jl` is a Julia package for a posteriori processing evaluation of time-resolved nonlinear spectroscopic signals from on-the-fly trajectory simulations using the quasi-classical DW approximation. The DW formalism is a robust method that allows the calculation and interpretation of femtosecond time-resolved third-order spectroscopic signals. In the framework of the DW approximation, the input to `WaveMixings.jl` are electronic energies and transition dipole moments computed along quasi-classical trajectories (SH, Ehrenfest, mapping Hamiltonian approach or similar algorithms) for the simulation of time-resolved nonlinear spectra. The implementation of the DW approximation in `WaveMixings.jl` provides a flexible and high-efficient tool for the computation and interpretation of femtosecond time-resolved spectra.

```julia
1  # WaveMixings parameters of the signal 2D
2  pump_frequ = make_pump_freq(5.2)
3  pulse_time = make_pulse_duration(0.1)
4  excit_freq = make_probe_freq(1.5, 6.0, 200)
5  # Read file with data from SH calculation
6  tmp_GS = extract_data_gs("001_good_gsb.log")
7  # Calculate the rephasing doorway function
8  # for the non-rephasing doorway function use gsb_door_nr2d
9  gsb_door_r2d(pump_frequ, pulse_time, excit_freq, tmp_GS)
10 # Calculate the window function
11 gsb_wind_2d(pump_frequ, pulse_time, excit_freq, tmp_GS, 121)
12 # Create the Doorway-window and store the result
13 # in the "spectral_signal_2d" field
14 make2D_door_window(tmp_GS)
```

Listing 2: Julia code example demonstrating the calculation of the GSB contribution to the 2D TA signal. The codes for the SE and ESA signals are similar.

In this article, we illustrated the application of `WaveMixings.jl` by presenting the simulated TA PP spectra for pyrazine, a prototypical system for ultrafast excited-state dynamics driven by conical intersections. The demonstration includes several nonlinear signals: the integral TA PP signal (Section VI A), the dispersed TA PP signal (Section VI B), the 2-dimensional TA signal (Section VI D), and time-resolved fluorescence (Section VI C). The package relies solely on standard Julia libraries and manages all underlying matrix operations automatically, streamlining its deployment and ensuring high performance. Additionally, `WaveMixings.jl` offers built-in tools and functions for simplifying file parsing, signal post-processing, and spectra plotting. Simulations of spectra with `WaveMixings.jl` require minimal user-specified input data, reducing the likelihood of errors.

The relevance of `WaveMixings.jl` for experimental and computational ultrafast time-resolved spectroscopy can be characterized as follows. First, `WaveMixings.jl` can serve as a model tool to simulate measured spectroscopic signals. This capability enables users to compare simulations with experimental data, if available, providing deeper insights into the underlying photophysics or photochemistry. Second, `WaveMixings.jl` can be used as a tool for designing spectroscopic experiments. In this role, it represents a valuable resource for planning experimental settings, such as pulse durations or the range of pulse carrier frequencies. Third, for educational purposes, `WaveMixings.jl` offers an effective tool of teaching linear and nonlinear spectroscopy concepts by utilizing computationally generated data. Students can visually grasp the fundamental concepts more easily, leading to a deeper understanding. Fourth, as an environment for the development of computational methods, `WaveMixings.jl` provides a straightforward and flexible platform for implementing the simulation of novel spectroscopic signals within the framework of the DW approximation.

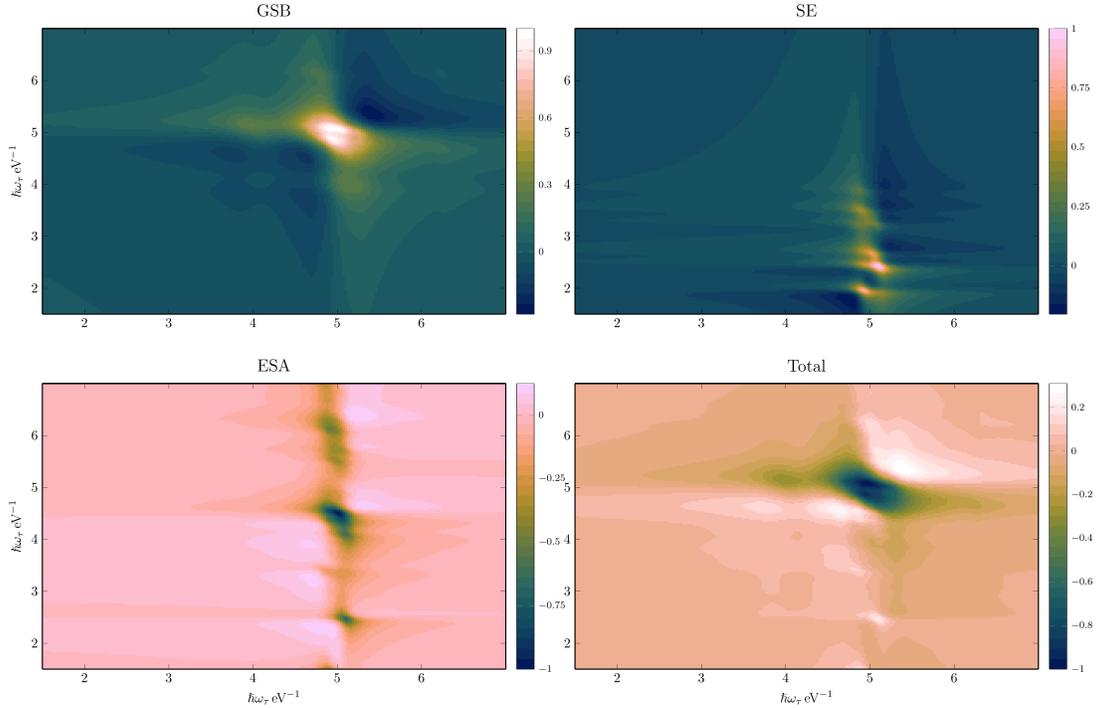

Figure 5: Pyrazine 2D TA signals (GSB, SE, and ESA) and their total contribution at waiting time 60 fs. `WaveMixings.jl` input values are: pulse duration ($\tau$) = 0.1 fs, pump frequency ($\omega_{\text{pu}}$) = 5.2 eV, 1st and 2nd excitation frequency ($\omega_{\text{pr}}$) 200 data points from 1.5 eV to 7.0 eV.

`WaveMixings.jl` is a tool in development. In future versions, more functionalities may be available. Current developments aim at adding alternative spectroscopic techniques to the portfolio, such as population-detected four-wave-mixing (FWM) signals, polarization-resolved signals, or six-wave-mixing signals. Additional objectives include enhancing plotting features, expanding the list of recognized software outputs, include anisotropies, support X-ray TA PP simulations, and providing more post-processing tools.

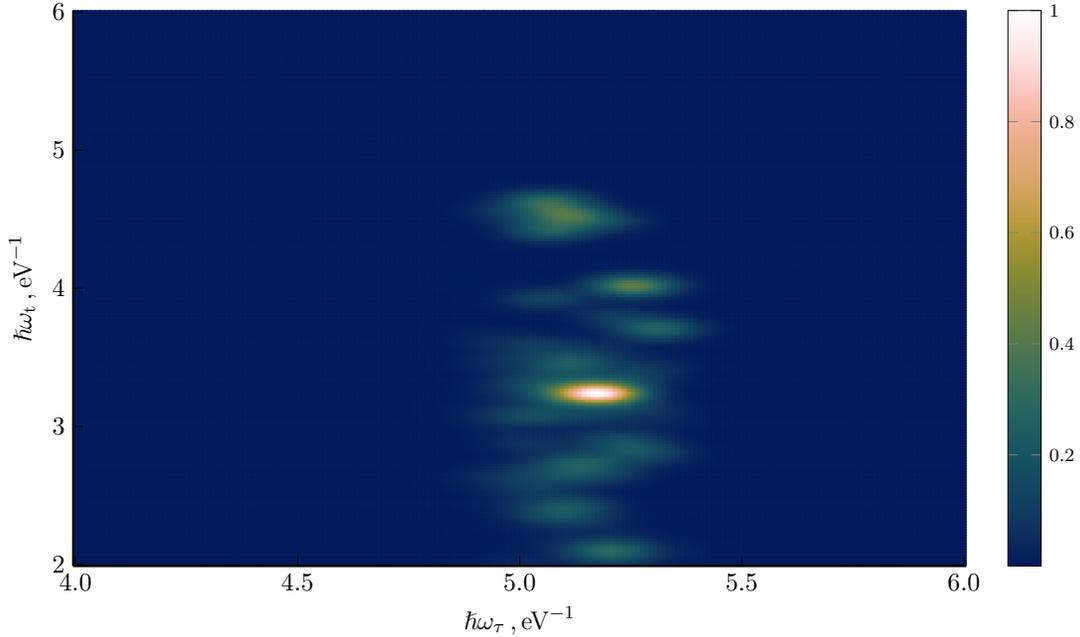

Figure 6: Normalized 2D-FLEX signal for pyrazine as a function of the excitation frequency ($\omega_\tau$) and the detection frequency ($\omega_t$) a waiting time of $60\,\text{fs}$. `WaveMixings.jl` input values are: pump frequency $\omega_{\text{pu}} = 5.2\,\text{eV}$, pump pulse duration $\tau_{\text{pu}} = 5\,\text{fs}$, and gate pulse duration $\tau_t = 30\,\text{fs}$.

**ACKNOWLEDGMENTS**

We thank Dr. Nađa Došlić and Dr. Marin Sapunar for their long-term support that made possible most simulations of spectra. This work is supported by the the National Natural Science Foundation of China (No. 22373028). Z.L. acknowledges support from the National Natural Science Foundation of China (No. 22333003, 22361132528). S.V.P. acknowledges support from the Key Research Project of Zhejiang Lab (No. 2021PE0AC02) and from the National Natural Science Foundation of China (No. W2433024). L.P.C. acknowledges support from the Key Research Project of Zhejiang Lab (No. 2021PE0AC02).

# REFERENCES

[1] O. Travnikova, T. Piteša, A. Ponzi, M. Sapunar, R. J. Squibb, R. Richter, P. Finetti, M. Di Fraia, A. De Fanis, and N. *et al.* Mahne, 'Photochemical ring-opening reaction of 1,3-cyclohexadiene: Identifying the true reactive state,' J. Am. Chem. Soc. **144**, 21878–21886 (2022).

[2] K. S. Zinchenko, F. Ardana-Lamas, I. Seidu, S. P. Neville, J. van der Veen, V. U. Lanfaloni, M. S. Schuurman, and H. J. Wörner, 'Sub-7-fs conical-intersection dynamics probed at the carbon K-edge,' Science **371**, 489–494 (2021).

[3] S. Karashima, A. Humeniuk, and T. Suzuki, 'Vibrational motions in ultrafast electronic relaxation of pyrazine,' J. Am. Chem. Soc. **146**, 11067–11071 (2024).

[4] S. Mukamel, *Principles of Nonlinear Optical Spectroscopy* (Oxford University Press, New York, 1995).

[5] M. Gelin, L. Chen, and W. Domcke, 'Equation-of-motion methods for the calculation of femtosecond time-resolved 4-wave-mixing and N-wave mixing signals,' Chem. Rev. **122**, 17339–17396 (2022).

[6] M. Klessinger and J. Michl, *Excited States and Photochemistry of Organic Molecules* (VCH Publishers, New York, 1995).

[7] W. Domcke, D. Yarkony, H. Köppel, and . (Eds.), *Conical intersections: Electronic structure, dynamics & spectroscopy*, Advanced series in physical chemistry (World Scientific, 2004).

[8] W. Domcke, D. Yarkony, H. Köppel, and . (Eds.), *Conical Intersections: Theory, Computation and Experiment*, Advanced series in physical chemistry (World Scientific, 2011).

[9] M. H. Beck, G. A. Jäckle, G. A. Worth, and H.-D. Meyer, 'The multiconfiguration time-dependent Hartree (MCTDH) method: a highly efficient algorithm for propagating wavepackets,' Phys Rep. **324**, 1–105 (2000).

[10] U. Manthe, 'Wavepacket dynamics and the multi-configurational time-dependent hartree approach,' J. Phys. Cond. Matt. **29**, 253001 (2017).

[11] I. V. Oseledets and E. E. Tyrtyshnikov, 'Breaking the curse of dimensionality and or how to use SVD in many dimensions,' SIAM J. Sci. Comput. **31**, 3744–3759 (2009).

[12] Y. Kurashige, 'Matrix product state formulation of the multiconfiguration time-dependent hartree theory,' J. Chem. Phys. **149**, 194114 (2018).

[13] R. Borrelli and S. Dolgov, 'Expanding the range of hierarchical equations of motion by tensor-train implementation,' J. Phys. Chem. A **125**, 5397–5407 (2021).

[14] T. G. Kolda and T. W. Bader, 'Tensor decompositions and applications,' SIAM Rev. **51**, 455–500 (2009).

[15] I. V. Oseledets, 'Tensor-train decomposition,' SIAM J. Sci. Comput. **33**, 2295–2317 (2011).

[16] G. Stock and M. Thoss, 'Classical description of nonadiabatic quantum dynamics,' Adv. Chem. Phys. **131**, 243–375 (2005).

[17] R. Crespo-Otero and M. Barbatti, 'Recent advances and perspectives on nonadiabatic mixed quantum-classical dynamics,' Chem. Rev. **118**, 7026–7068 (2018).

[18] T. R. Nelson, A. J. White, J. A. Bjorgaard, A. E. Sifain, Y. Zhang, B. Nebgen, S. Fernandez-Alberti, D. Mozyrsky, A. E. Roitberg, and S. Tretiak, 'Non-adiabatic excited-state molecular dynamics: theory and applications for modelling photophysics in extended molecular materials,' Chem. Rev. **120**, 2215–2287 (2020).

[19] D. Avagliano, M. Bonfanti, A. Nenov, and M. Garavelli, 'Automatized protocol and interface to simulate QM/MM timeresolved transient absorption at TDDFT level with COBRAMM,' J. Comput. Chem. **43**, 1641–1655 (2022).

[20] F. Segatta, A. Nenov, D. R. Nascimento, N. Govind, S. Mukamel, and M. Garavelli, 'iSPECTRON: A simulation interface for linear and nonlinear spectra with abinitio quantum chemistry software,' J. Comput. Chem. **42**, 644–659 (2021).

[21] F. Segatta, D. A. Ruiz, F. Aleotti, M. Yaghoubi, S. Mukamel, M. Garavelli, F. Santoro, and A. Nenov, 'Nonlinear molecular electronic spectroscopy via MCTDH quantum dynamics: From exact to approximate expressions,' J. Chem. Theory Comput. **19**, 2075–2091 (2023).

[22] A. Loreti, V. M. Freixas, D. Avagliano, F. Segatta, H. Song, S. Tretiak, S. Mukamel, M. Garavelli, N. Govind, and A. Nenov, 'WFOT: A wave function overlap tool between single- and multi-reference electronic structure methods for spectroscopy simulation,' J. Chem. Theory Comput. **20**, 4804–4819 (2024).

[23] N. Tancogne-Dejean, M. J. T. Oliveira, X. Andrade, H. Appel, C. H. Borca, G. Le Breton, F. Buchholz, A. Castro, S. Corni, A. A. Correa, U. De Giovannini, A. Delgado, F. G. Eich, J. Flick, G. Gil, A. Gomez, N. Helbig, H. Hübener, R. Jestädt, J. Jornet-Somoza, A. H. Larsen, I. V. Lebedeva, M. Lüders, M. A. L. Marques, S. T. Ohlmann, S. Pipolo, M. Rampp, C. A. Rozzi, D. A. Strubbe, S. A. Sato, C. Schäfer, I. Theophilou, A. Welden, and A. Rubio, 'Octopus, a computational framework for exploring light-driven phenomena and quantum

dynamics in extended and finite systems,' J. Chem. Phys. **152** (2020), 10.1063/1.5142502.

[24] S. Gozem and A. I. Krylov, 'The ezspectra suite: An easytouse toolkit for spectroscopy modeling,' WIREs Comput. Mol. Sci. **12** (2021), 10.1002/wcms.1546.

[25] T. Kenneweg, S. Mueller, T. Brixner, and W. Pfeiffer, 'QDT  A Matlab toolbox for the simulation of coupled quantum systems and coherent multidimensional spectroscopy,' Comput. Phys. Commun. **296**, 109031 (2024).

[26] Y. J. Yan, L. E. Fried, and S. Mukamel, 'Ultrafast pump-probe spectroscopy: femtosecond dynamics in liouville space,' J. Phys. Chem. **93**, 8149–8162 (1989).

[27] Y. J. Yan and S. Mukamel, 'Femtosecond pump-probe spectroscopy of polyatomic molecules in condensed phases,' Phys. Rev. A **41**, 6485–6504 (1990).

[28] L. E. Fried and S. Mukamel, 'A classical theory of pump-probe photodissociation for arbitrary pulse durations,' J. Chem. Phys. **93**, 3063–3071 (1990).

[29] M. F. Gelin, A. V. Pisliakov, and W. Domcke, 'Time and frequency gated spontaneous emission as a tool for studying vibrational dynamics in the excited state,' Phys. Rev. A **65**, 062507 (2002).

[30] A. V. Pisliakov, M. F. Gelin, and W. Domcke, 'Detection of electronic and vibrational coherence effects in electron-transfer systems by femtosecond time-resolved fluorescence spectroscopy: theoretical aspects,' J. Phys. Chem. A **107**, 2657–2666 (2003).

[31] A. Okada, V. Chernyak, and S. Mukamel, 'Solvent reorganization in long-range electron transfer: density matrix approach,' J. Phys. Chem. A **102**, 1241–1251 (1998).

[32] V. Chernyak, T. Minami, and S. Mukamel, 'Exciton transport in molecular aggregates probed by time and frequency gated optical spectroscopy,' J. Chem. Phys **112**, 7953–7963 (2000).

[33] J. C. Kirkwood, C. Scheurer, V. Chernyak, and S. Mukamel, 'Simulations of energy funneling and time- and frequency-gated fluorescence in dendrimers,' J. Chem. Phys. **114**, 2419–2429 (2001).

[34] R. Tempelaar, F. C. Spano, J. Knoester, and T. L. C. Jansen, 'Mapping the evolution of spatial exciton coherence through time-resolved fluorescence,' J. Phys. Chem. Lett. **5**, 1505–1510 (2014).

[35] P. C. Arpin and D. B. Turner, 'Signatures of vibrational and electronic quantum beats in femtosecond coherence spectra,' J. Phys. Chem. A **125**, 2425–2435 (2021).

[36] M. F. Gelin and D. S. Kosov, 'Doorway-window description of sequential three-pulse photon

echo signals,' Chem. Phys. **347**, 177–184 (2008).

[37] M. F. Gelin and W. Domcke, 'Alternative view of two-dimensional spectroscopy,' J. Chem. Phys. **144**, 194104 (2016).

[38] V. I. Novoderezhkin, T. A. Cohen Stuart, and R. van Grondelle, 'Dynamics of exciton relaxation in LH2 antenna probed by multipulse nonlinear spectroscopy,' J. Phys. Chem. A **115**, 3834–3844 (2011).

[39] M. F. Gelin, D. Egorova, and W. Domcke, 'Manipulating electronic couplings and nonadiabatic nuclear dynamics with strong laser pulses,' J. Chem. Phys. **131**, 124505 (2009).

[40] M. F. Gelin, D. Egorova, and W. Domcke, 'Optical N-wave-mixing spectroscopy with strong and temporally well-separated pulses: The doorway-window representation,' J. Phys. Chem. B **115**, 5648–5658 (2011).

[41] M. F. Gelin, D. Egorova, and W. Domcke, 'Strong-pump strong-probe spectroscopy: Effects of higher excited electronic states,' Phys. Chem. Chem. Phys. **15**, 8119–8131 (2013).

[42] L. Chen, E. Palacino-González, M. F. Gelin, and W. Domcke, 'Nonperturbative response functions: A tool for the interpretation of four-wave-mixing signals beyond third order,' J. Chem. Phys. **147**, 234104 (2017).

[43] G. Bressan and J. J. van Thor, 'Theory of two-dimensional spectroscopy with intense laser fields,' J. Chem. Phys. **154**, 244111 (2021).

[44] V. Novoderezhkin, R. Monshouwer, and R. van Grondelle, 'Electronic and vibrational coherence in the core light-harvesting antenna of Rhodopseudomonas viridis,' J. Phys. Chem. B **104**, 12056–12071 (2000).

[45] V. I. Novoderezhkin, M. A. Palacios, H. van Amerongen, and R. van Grondelle, 'Energy-transfer dynamics in the LHCII complex of higher plants: Modified Redfield approach,' J. Phys. Chem. B **108**, 10363–10375 (2004).

[46] B. Balzer and G. Stock, 'Transient spectral features of a cis-trans photoreaction in the condensed phase: a model study,' J. Phys. Chem. A **108**, 6464–6473 (2004).

[47] W. M. Zhang, T. Meier, V. Chernyak, and S. Mukamel, 'Exciton-migration and three-pulse femtosecond optical spectroscopies of photosynthetic antenna complexes,' J. Chem. Phys **108**, 7763–7774 (1998).

[48] S. Mukamel and D. Abramavicius, 'Many-body approaches for simulating coherent nonlinear spectroscopies of electronic and vibrational excitons,' Chem. Rev. **104**, 2073–2098 (2004).

[49] D. Abramavicius, B. Palmieri, D. V. Voronine, F. Šanda, and S. Mukamel, 'Coherent multidimensional optical spectroscopy of excitons in molecular aggregates and quasiparticle *vs* supermolecule perspective,' Chem. Rev. **109**, 2350–2408 (2009).

[50] M. F. Gelin, A. V. Pisliakov, D. Egorova, and W. A. Domcke, 'Simple model for the calculation of nonlinear optical response functions and femtosecond time-resolved spectra,' J. Chem. Phys. **118**, 5287–5301 (2003).

[51] T. Hasegawa and Y. Tanimura, 'Nonequilibrium molecular dynamics simulations with a backward-forward trajectories sampling for multidimensional infrared spectroscopy of molecular vibrational modes,' J. Chem. Phys. **128**, 064511 (2008).

[52] G. Stock, 'Classical description of nonadiabatic photoisomerization processes and their real-time detection via femtosecond spectroscopy,' J. Chem. Phys. **103**, 10015–10029 (1995).

[53] I. Uspenskiy, B. Strodel, and G. Stock, 'Classical calculation of transient absorption spectra monitoring ultrafast electron transfer processes,' J. Chem. Theory Comput. **2**, 1605–1617 (2006).

[54] B. J. Schwartz and P. J. Rossky, 'Pump-probe spectroscopy of the hydrated electron: A quantum molecular dynamics simulation,' J. Chem. Phys. **101**, 6917–6926 (1994).

[55] Z. Li, J.-Y. Fang, and C. C. Martens, 'Simulation of ultrafast dynamics and pump-probe spectroscopy using classical trajectories,' J. Chem. Phys. **104**, 6919–6929 (1996).

[56] J. Che, W. Zhang, and Y. J. A. Yan, 'Classical time-frequency theory of transient absorption spectroscopy,' J. Chem. Phys. **106**, 6947–6956 (1997).

[57] L. W. Ungar and J. A. Cina, 'Short-time fluorescence Stokes shift dynamics,' Adv. Chem. Phys. **100**, 171–228 (1997).

[58] V. A. Ermoshin and V. Engel, 'Femtosecond pump-probe fluorescence signals from classical trajectories: Comparison with wave-packet calculations,' Eur. Phys. J. D **15**, 413–422 (2001).

[59] A. Heidenreich and J. Jortner, 'Pump-probe spectroscopy of ultrafast structural relaxation of electronically excited rare gas heteroclusters,' J. Electr. Spectrosc. Relat. Phen. **106**, 187–197 (2000).

[60] V. Bonačić-Koutecky and R. Mitrić, 'Theoretical exploration of ultrafast dynamics in atomic clusters: Analysis and control,' Chem. Rev. **105**, 11–65 (2005).

[61] R. Mitrić, V. Bonačić-Koutecký, J. Pittner, and H. Lischka, 'Ab initio nonadiabatic dynamics study of ultrafast radiationless decay over conical intersections illustrated on the

Na$_3$F cluster,' Chem. Phys. **125**, 024303 (2006).

[62]R. Mitrić, J. Petersen, and V. Bonačić-Koutecký, 'Multistate nonadiabatic dynamics "on the fly" in complex systems and its control by laser fields,' in *Conical Intersections: Theory, Computation and Experiment*, edited by W. Domcke, D. R. Yarkony, and H. Köppel (World Scientific, Singapore, 2011) pp. 497–568.

[63]M. F. Gelin, X. Huang, W. Xie, L. Chen, N. Došlić, and W. Domcke, 'Ab initio surface hopping simulation of femtosecond transient-absorption pump-probe signals of nonadiabatic excited-state dynamics using the doorway-window representation,' J. Chem. Theory. Comput. **17**, 2394–2408 (2021).

[64]X. Huang, W. Xie, N. Došlić, M. F. Gelin, and W. Domcke, 'Ab initio quasiclassical simulation of femtosecond time-resolved two-dimensional electronic spectra of pyrazine,' J. Phys. Chem. Lett. **12**, 11736–11744 (2021).

[65]C. Xu, C. Lin, J. Peng, J. Zhang, S. Lin, F. L. Gu, M. F. Gelin, and Z. Lan, 'On-the-fly simulation of time-resolved fluorescence spectra and anisotropy,' J. Chem. Phys. **160** (2024).

[66]S. V. Pios, M. F. Gelin, L. Vasquez, J. Hauer, and L. Chen, 'On-the-fly simulation of two-dimensional fluorescence-excitation spectra,' J. Phys. Chem. Lett. **15**, 8728–8735 (2024).

[67]H. Guan, K. Sun, L. Vasquez, L. Chen, S. V. Pios, Z. Lan, and M. F. Gelin, 'Quasiclassical Doorway-Window simulation of femtosecond transient-absorption pump-probe signals beyong the weak-field limit,' J. Chem. Theory Comput. **21**, 7561–7575.

[68]K. Sun, L. Vasquez, R. Borrelli, L. Chen, Y. Zhao, and M. F. Gelin, 'Interconnection between polarization-detected and population-detected signals: Theoretical results and ab initio simulations,' J. Chem. Theory Comput. **20**, 7560–7573 (2024).

[69]M. F. Gelin, Z. Lan, N. Došlić, and W. Domcke, 'Computation of timeresolved nonlinear electronic spectra from classical trajectories,' WIREs Comput. Mol. Sci. **15** (2025), 10.1002/wcms.70012.

[70]C. Xu, K. Lin, D. Hu, F. L. Gu, M. F. Gelin, and Z. Lan, 'Ultrafast internal conversion dynamics through the on-the-fly simulation of transient absorption pumpprobe spectra with different electronic structure methods,' J. Phys. Chem. Lett. **13**, 661–668 (2022).

[71]L. Vasquez, 'WaveMixings.jl,' (2023), https://codeberg.org/apolionl/WaveMixings.jl.

[72]L. Vasquez, M. Gelin, W. Domcke, Z. Lan, S. V. Pios, and L. Chen, 'Transient absorption pump-probe simulation in-/outputs – WaveMixings: Julia package for performing

on-the-fly time-resolved nonlinear electronic spectra from quasi-classical trajectories,' https://doi.org/10.5281/zenodo.15598121.

[73] M. Hillery, R. F. O'Connel, M. O. Scully, and E. P. Wigner, 'Distribution functions in physics: fundamentals,' Phys. Rep. **106**, 121–167 (1984).

[74] C. Xu, C. Lin, J. Peng, J. Zhang, S. Lin, F. L. Gu, M. F. Gelin, and Z. Lan, 'On-the-fly simulation of time-resolved fluorescence spectra and anisotropy,' J. Chem. Phys. **160**, 104109 (2024).

[75] D. Hu, J. Peng, L. Chen, M. F. Gelin, and Z. Lan, 'Spectral fingerprint of excited-state energy transfer in dendrimers through polarization-sensitive transient-absorption pump–probe signals: On-the-fly nonadiabatic dynamics simulations,' J. Phys. Chem. Lett. **12**, 9710–9719 (2021).

[76] R. Perez-Castillo, V. M. Freixas, S. Mukamel, A. Martinez-Mesa, L. Uranga-Pina, S. Tretiak, M. F. Gelin, and S. Fernandez-Alberti, 'Transient-absorption spectroscopy of dendrimers via nonadiabatic excited-state dynamics simulations,' Chem. Sci. **15**, 13250–13261 (2024).

[77] J. Yang, M. F. Gelin, L. Chen, F. Šanda, E. Thyrhaug, and J. Hauer, 'Two-dimensional fluorescence excitation spectroscopy: A novel technique for monitoring excited-state photophysics of molecular species with high time and frequency resolution,' J. Chem. Phys. **159**, 74201 (2023).

[78] J. Novak, M. Mališ, A. Prlj, I. Ljubić, O. Kühn, and N. Došlić, 'Photoinduced dynamics of formic acid monomers and dimers: The role of the double hydrogen bond,' J. Phys. Chem. A **116**, 11467–11475 (2012).

[79] M. Sapunar, T. Piteša, D. Davidović, and N. Došlić, 'Highly efficient algorithms for cis type excited state wave function overlaps,' J. Chem. Theory Comput. **15**, 3461–3469 (2019).

[80] M. Sapunar, 'ZagHop,' https://github.com/marin-sapunar/ZagHop.